\documentclass[pre,floatfix,twocolumn,10pt]{revtex4-1}
\usepackage{graphicx}
\usepackage{amssymb}
\usepackage{dcolumn}
\usepackage{bm}

\begin{document}

\author{I. Fern\'andez Aguirre and E. A. Jagla} 
\affiliation{Comisi\'on Nacional de Energ\'{\i}a At\'omica, Instituto Balseiro (UNCu), and CONICET\\
Centro At\'omico Bariloche, (8400) Bariloche, Argentina}

\title{On the critical exponents of the yielding transition of amorphous solids}

\begin{abstract} 

We investigate numerically the yielding transition of a two dimensional model amorphous solid under external shear. 
We use a scalar model in terms of values of the total local strain, that we derive from the full (tensorial) description of the elastic interactions in the system, in which
plastic deformations are accounted for by introducing a stochastic ``plastic disorder" potential.
This scalar model is seen to be equivalent to a collection of Prandtl-Tomlinson particles,  which are coupled through an Eshelby quadrupolar kernel. Numerical simulations of this scalar model reveal that the strain rate vs stress curve, close to the critical stress, is of the form $\dot\gamma\sim (\sigma-\sigma_c)^\beta$. Remarkably, we find that the value of $\beta$ depends on details of the microscopic plastic potential used, confirming and giving additional support to results previously obtained with the full tensorial model. %\cite{Jagla_Yiel}. 
To rationalize this result,  we argue that the Eshelby interaction in the scalar model can be treated to a good approximation in a sort of ``dynamical" mean field, which corresponds to a Prandtl-Tomlinson particle that is driven by the applied strain rate in the presence of a stochastic noise generated by all other particles. The dynamics of this Prandtl-Tomlinson particle displays different values of the $\beta$ exponent depending on the analytical properties of the microscopic potential, thus giving support to the results of the numerical simulations.
Moreover, we find that other critical exponents that depend on details of the dynamics show also a dependence with the form of the disorder,  while static exponents are independent of the details of the disorder.
Finally, we show how our scalar model relates to other elastoplastic models and to the widely used mean field version known as the H\'ebraud-Lequeux model.

\end{abstract}

\maketitle

\section{Introduction}

Amorphous solid materials are ubiquitous in every day life, and of great practical importance in many industrial processes \cite{Ferrero}. They consist of a collection of elementary units that accommodate in space without a well defined ordering (contrary to what happens with crystals). The nature of the elementary units that form the material may span a wide range, roughly from $\sim 0.1~nm$ to $\sim 1~ m$\cite{Ferrero}. The fact that in many cases these units are not microscopic leads to the fact that thermal fluctuations may be negligible in explaining the mechanical properties of these materials, which are then termed ``athermal".

In recent years, there has been an increasing effort aimed at elucidating the mechanical properties of amorphous solids. 
One main piece of the phenomenology of amorphous materials is the existence of a yielding transition: In the absence of appreciable thermal activation effect, the material remains rigid if the applied stress is below some threshold, and it flows continuously if this threshold is exceeded. The properties of the material around this critical stress $\sigma_c$, or yield point, has attracted much attention. It is experimentally found  \cite{FC_1,FC_2}  that the strain rate in the system $\dot\gamma$ as a function of stress excess $\sigma-\sigma_c$ follows in many cases a power law behavior of the form $\dot\gamma\sim (\sigma-\sigma_c)^\beta$.  The flow exponent $\beta$ is an important parameter characterizing the problem.
Other important critical exponents emerge when one considers the nature of the dynamics close to the transition. This dynamics proceeds through abrupt rearrangements in the system \cite{SZ_Exp_1, SZ_Exp_2, SZ_Exp_3, SZ_Exp_4, SZ_Exp_5, SZ_Exp_6, SZ_Exp_7, SZ_Exp_8, SZ_Exp_9, SZ_Exp_10, SZ_Exp_11, SZ_Num_1, SZ_Num_2, SZ_Num_3, SZ_Num_4, SZ_Num_5, SZ_Num_6, SZ_Num_7, SZ_Num_8, SZ_Num_9, SZ_Num_10, SZ_Num_11}, which share many features with the avalanches observed in the related model of depinning of an elastic interface\cite{fisher,kardar}. This allowed to use in the analysis of the yielding transition the previously developed tools used in the depinning problem.\cite{Rosso_PNAS}
In particular one can define for yielding additional critical exponents associated to the statistics of avalanches close to the transition.

Since the universal aspects of the depinning transitions are well known, the issue of the universality of the yielding transition has attracted much interest.  In particular, are the values of the critical exponents independent of details of the model and only dependent on some very general characteristics as dimensionality, for instance? And if this is not the case, what are the system features that determine the differences?

In this work we argue that there are differences in the values of some critical exponents in the yielding problem, related to the form of the plastic yielding potential that is used to model the plastic rearrangements in the system. This is an interesting finding since it does not occur in the (short range) depinning problem. 
We find evidence that this result is related to the long range nature of the elastic interactions in the yielding problem, which leads to a sort of effective ``dynamical mean field'' description. In fact, in mean field depinning the same dependence of exponent $\beta$ on the form of the pinning potential is well known.\cite{fisher,Narayan}

The paper is organized as follows. In section II we present the model, which is a reduction to a scalar problem of a tensorial model of the yielding transition that has been presented previously \cite{Jagla_Yiel}. In section III the main numerical results are presented, showing the dependences of some critical exponents on the form of the plastic disorder potential. Section IV contains the arguments leading to a ``mean field like" description of the problem and then to the justification of the different values of the critical exponents found numerically. In section V we discuss to what extent the present model is comparable to the elastoplastic models
discussed in the literature. Finally, in Section VI we summarize and conclude.

\section{Model}

We motivate here the model in an heuristic way, emphasizing the physical ingredients it incorporates. 
In Ref. \cite{Jagla_Yiel} there is a derivation of the model from a full tensorial description of the elasticity of the material. In addition, in Appendix I, we present an alternative view in which a very similar model is deduced assuming the deformation field of the material is strictly one-dimensional under a single shear imposed deformation. 

The mesoscopic model we present describes the evolution of the system under a given external shear deformation of uniform symmetry. The goal is to predict the evolution of the corresponding local deformation $e(r)$ compatible with the externally applied load.
%Although the model disregards other deformation modes, they are implicitly c
The dynamics to be used is an over-damped dynamics in which the rate of change of $e(r)$ is equalled to an effective force acting at $r$. There are two main parts of this force. One is a local term encoding the internal dynamics of the element at $r$. This part is derived from a potential function $V_r(e)$. The form of $V_r(e)$ takes into account both the local elasticity of the material and also the possibility of different locally stable configurations: $V_r(e)$ has minima at a sequence of $e$ values, corresponding to equilibrium configurations. The transition between consecutive minima correspond to plastic events in the system. Around each minima $V_r(e)$ behaves quadratically, reflecting the elasticity at the actual configuration. 
In addition, there is a term in the evolution equation that reflects the elastic interaction between elements at different spatial positions. This term is written in terms of a kernel $G(r-r')$ which is usually referred to as the Eshelby kernel. The model reads (using a discrete spatial representation):

\begin{equation}
\eta \dot e_i=-\frac{dV_i(e_i)}{de_i}+\sum_{j} G_{ij}e_{j}+\sigma
\label{modelo}
\end{equation}
where $\sigma$ is the applied stress. From now on we will set the viscous damping coefficient $\eta$ to $\eta=1$.
$G_{ij}$ depends only on the distance between $i$ and $j$, and is more compactly described by its Fourier transform $G_{\bf q}$:
\begin{equation}
G_{\bf q}=-\frac{2\mu B(q_x^2-q_y^2)^2}{\mu q^4+2B q_x^2q_y^2}
\label{gdeq}
\end{equation}
and $G_{{\bf q}=0}=0$. $B$ and $\mu$ are the bulk and shear modulus of the material.
These two equations define the model completely.
Since $\sum_i G_{ij}\sim G_{{\bf q}=0}=0$, spatially averaging Eq. (\ref{modelo}) we obtain
\begin{equation}
\dot {\overline {e}}=-\overline{\frac{dV_i(e)}{de}}+\sigma,
\label{modelo2}
\end{equation}
that determines the instantaneous value of the deformation rate $\dot\gamma\equiv\dot {\overline {e}}$.
Alternatively, in an implementation that fixes the value of the deformation rate $\dot\gamma$, Eq. (\ref{modelo2}) defines the value of the instantaneous stress as
\begin{equation}
\sigma=\dot {\gamma}+\overline{\frac{dV_i(e)}{de}}
\label{modelo3}
\end{equation}

Note that in the present formalism there is a single quantity $e_i$ for each site describing the state of the system, and the separation 
between elastic strain and plastic strain usually done in elasto-plastic models is not made. We will come back to the relation with other elasto-plastic models later on.
%In the present model, the separation between elastic and plastic strain must eventually be searched for in the form of the potentials $V(e)$. In this respect, we have found recently an unexpected dependency of model properties with the form of $V(e)$ and we want to explore here deeply this dependency. 

\begin{figure}
	\includegraphics[width=7cm,clip=true]{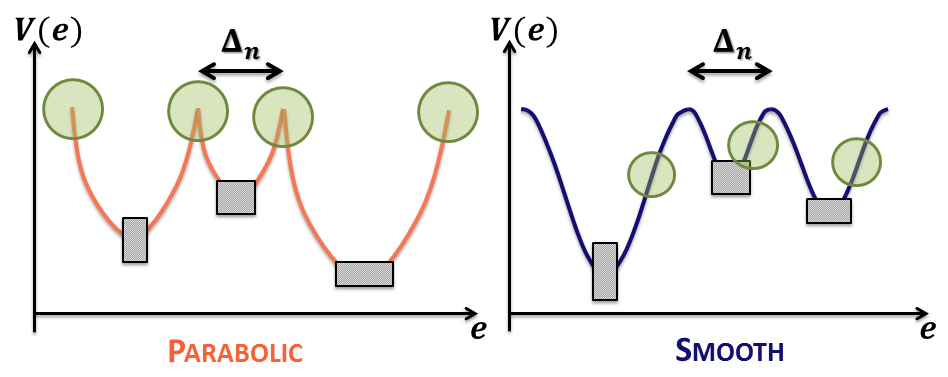}
	\caption{Schematic forms of the two plastic potentials used: parabolic (left) and smooth (right). Circles indicate the transition points (maximum force) under a global force pointing to the right. $\Delta_n$ is the well's width which is stochastically chosen from a flat distribution.
	\label{Fig:Pozos_de_Pot}
	}
\end{figure}

The $V(e)$ are stochastic potentials chosen to be uncorrelated among different spatial positions. The values of $e$ at which $V(e)$
has local minima correspond to locally stable configurations of the system. The form of $V(e)$ is quadratic around these minima to model an elastic material. To fully define the form of $V(e)$ we must specify how the wells corresponding to different minima are connected.
We consider two qualitatively different forms of the $V(e)$ potentials (see Fig. \ref{Fig:Pozos_de_Pot}). 
In the first case the wells are connected sharply, at points in which $dV(e)/de$ has jumps. In the second case the connection is made smoothly. 

In concrete, to define a potential $V(e)$, the $e$ axis is divided in intervals $[a_n,a_{n+1}]$  ($n$ integer), in such a way that 
$\Delta_n\equiv{a_{n+1}-a_{n}}$ is stochastically chosen from a flat distribution between $\Delta_{min}=2$ and $\Delta_{max}=4$  (we have checked that the use of an exponential distribution does not affect the results). 

The intervals are centered at $\overline{a_n}\equiv(a_{n+1}+a_{n})/2$. At each interval $n$, $V(n)$ is defined as 
\begin{equation}
{V}_n(e)=\frac{1}{2}\left[(e-\overline{a_n})^2-{\Delta_n}^2\right]
\label{Ec:V_Parabolic}
\end{equation}
for the case in Fig. \ref{Fig:Pozos_de_Pot}(a), and as
\begin{equation}
{V}_n(e)=-5\left(\frac{\Delta_n}{2\pi}\right)^2\left[1+\cos{\left(\frac{2\pi(e-\overline{a_n})}{\Delta_n}\right)}\right]
\label{Ec:V_Suave}
\end{equation}
for the case in Fig. \ref{Fig:Pozos_de_Pot}(b). The first case will be referred to as the ``parabolic" potential, and the second case as the ``smooth" potential.
The qualitative main difference between the two cases concerns the behavior at the transition points between different wells.
These are the point of maximum force, which are indicated as green dots in Fig. \ref{Fig:Pozos_de_Pot}. 
In the parabolic case these points coincide with the potential maxima, where there is a discontinuity in the force.
In the smooth potential case, the maximum force occurs at points where the curvature of the potential changes sign continuously.

\section{Results of numerical simulations}

In this Section we present results of simulations of the model, to elucidate the effect of the form of the potential on the critical exponents of the transition and the avalanche statistics. We set units such that $B=1$, and work in the case $\mu=B$.

\begin{figure}[h!]  
%	\centering  
%	\begin{subfigure}{.5\textwidth}  
%		\centering  
		\includegraphics[width=8cm,clip=true]{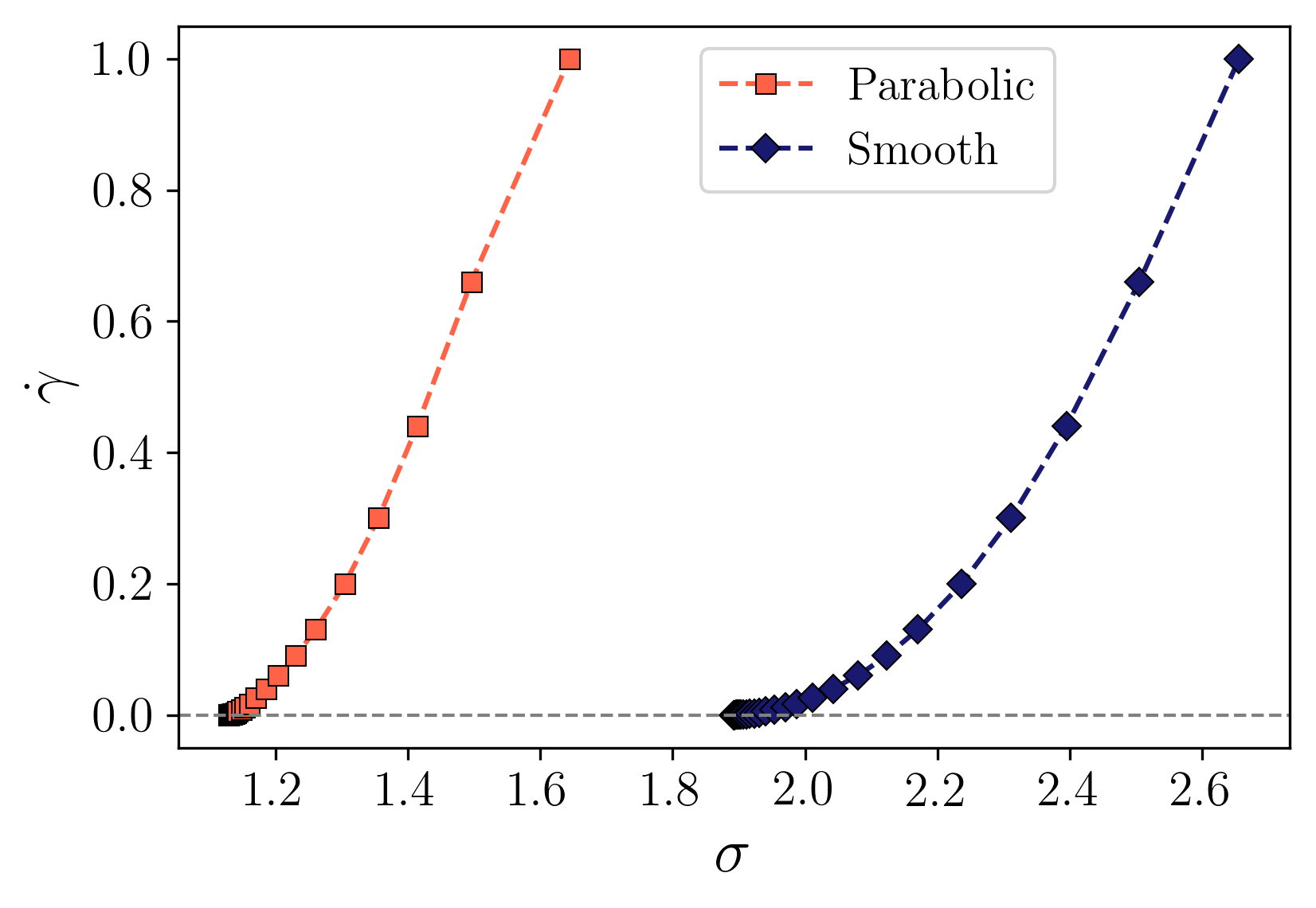}
		%\caption{Caption for image 1}   
%	\end{subfigure}   
	
%	\begin{subfigure}{.5\textwidth}  
%		\centering  
		\includegraphics[width=8cm,clip=true]{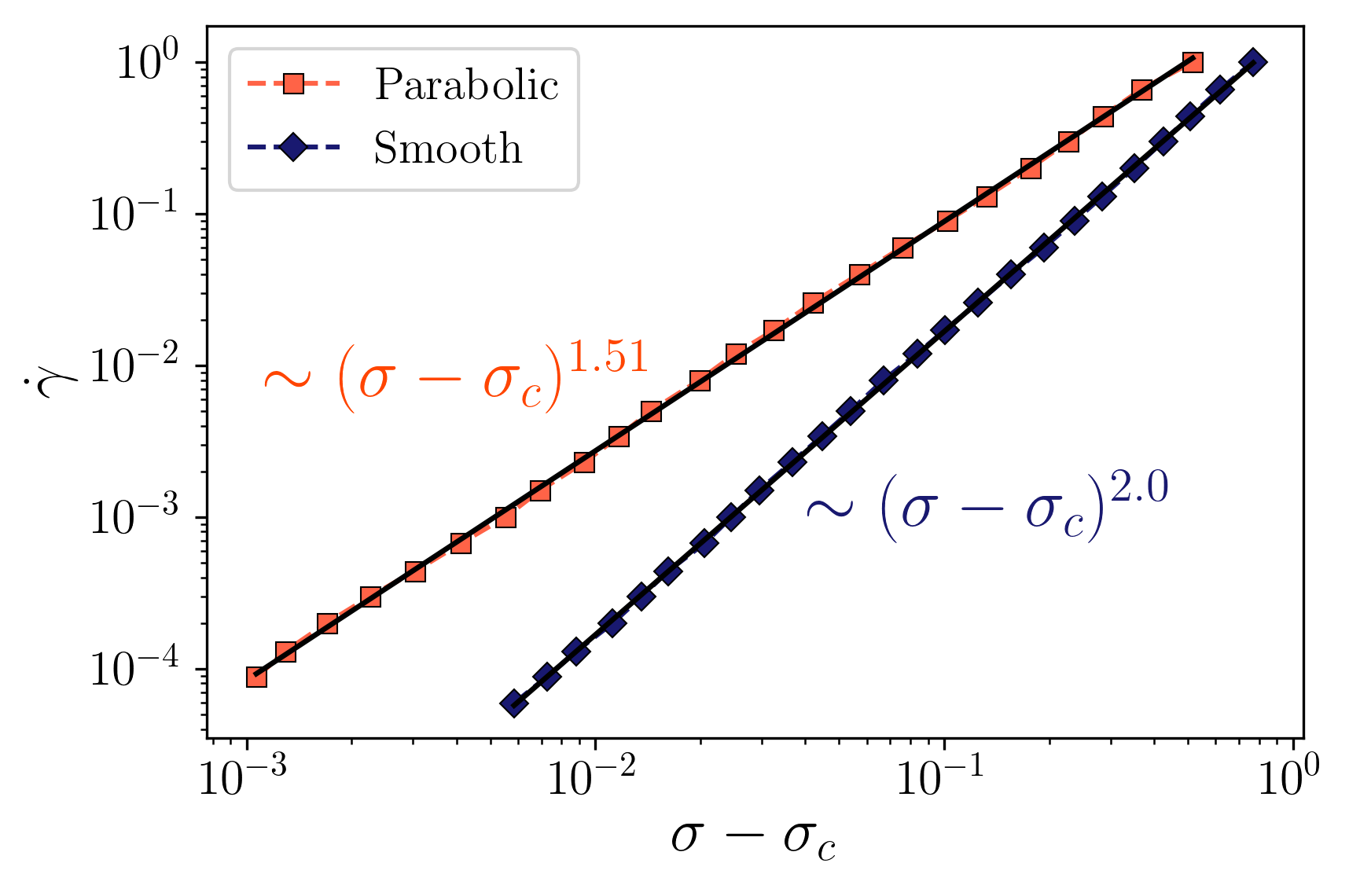}
		%\caption{Caption for image 2}   
%	\end{subfigure}  
	\caption{Strain rate vs. stress curves for smooth and parabolic potential. System size is $L=128$. Upper panel: Linear scale. Lower panel: Logarithmic scale with the value of $\sigma_c$ subtracted. Continuous black lines are the linear fits which provide  exponents indicated.
	\label{Fig:FLow}}  
\end{figure}

We focus first on the value of the flow exponent $\beta$. The value of $\beta$ can be measured straightforwardly by driving the system at a constant strain rate, and measuring the stress as it is defined in Eq. (\ref{modelo3}). As a result, we obtain the flow curves shown in the Figure \ref{Fig:FLow}. This graph displays clearly the existence of a critical stress and a monotonic growth for larger stress. The logarithmic plot in \ref{Fig:FLow}(b) clearly indicate that the values of $\beta$ are dependent on the form of the potential. We obtain $\beta_{p}\simeq1.51$ for parabolic potentials and  $\beta_{s}\simeq2.00$ for smooth potentials.

\begin{figure}[ht!]
	\includegraphics[width=8cm,clip=true]{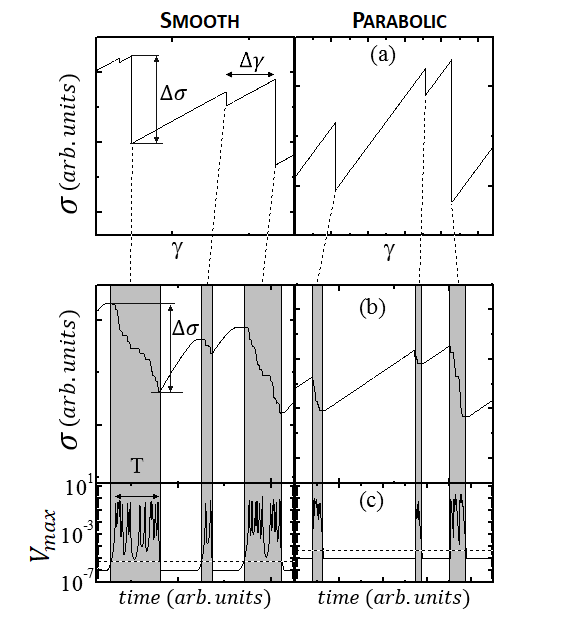}
	\caption{Examples of the evolution of stress in the system, under the quasi-static protocol described in the text. Right part corresponds to parabolic potentials, and left part to smooth potentials. In (a) we see the stress-strain plot, and in (b) the stress-time one. Strain rate is zero in the grey regions (when $V_{max}$, shown in panel (c), is larger than a threshold value highlighted with a dashed line), whereas it is a fixed small $\dot{\gamma}$ outside these periods. Each grey region corresponds to one avalanche. The size $S$ of each avalanche is obtained from the strain drop as $S= \overline{\Delta \sigma} L^2$. Avalanche duration $T$ 
is determined using a threshold criterion in $V_{max}$ (see text).
$\Delta \gamma$ corresponds to the strain increase that has to be applied after one avalanche to trigger a second one.
	\label{Fig:ava}}
\end{figure}

Motivated by this difference between the two kinds of potentials, we moved to study the exponents characterizing the avalanche dynamics. In order to calculate these quantities and to see in particular if they depend on the kind of potential used, we ran quasistatic simulations in the following way (see Figure \ref{Fig:ava}). In a simulation with a small $\dot{\gamma}$, the maximum value of $de/dt$ across the system is calculated: $V_{max}\equiv \max_i (de_i/dt)$ . This quantity stays lower than a well-chosen threshold as long the system is stable. However, when an avalanche is developing this quantity becomes order 1. When the avalanche finishes $V_{max}$ becomes very small again. In this way we can identify individual avalanches in the system.
It is important to point out that to obtain more precise results we stop the driving while an avalanche is taking place. This avoids spurious additional avalanche triggering by the driving.

As indicated in Fig. \ref{Fig:ava}, avalanche size $S$ is proportional to the global stress drop that the avalanche causes, and its duration $T$ is measured as the time between the first jump of any site from one minimum to another and the last one. 
%This quantity has an offset that can be fixed during the data analysis. 
Additionally, we also monitor the strain increases $\Delta \gamma$ that have to be applied after one avalanche to trigger a second one.

\begin{figure}[h!]  
%	\centering  
%	\begin{subfigure}{.5\textwidth}  
%		\centering  
		\includegraphics[width=8cm,clip=true]{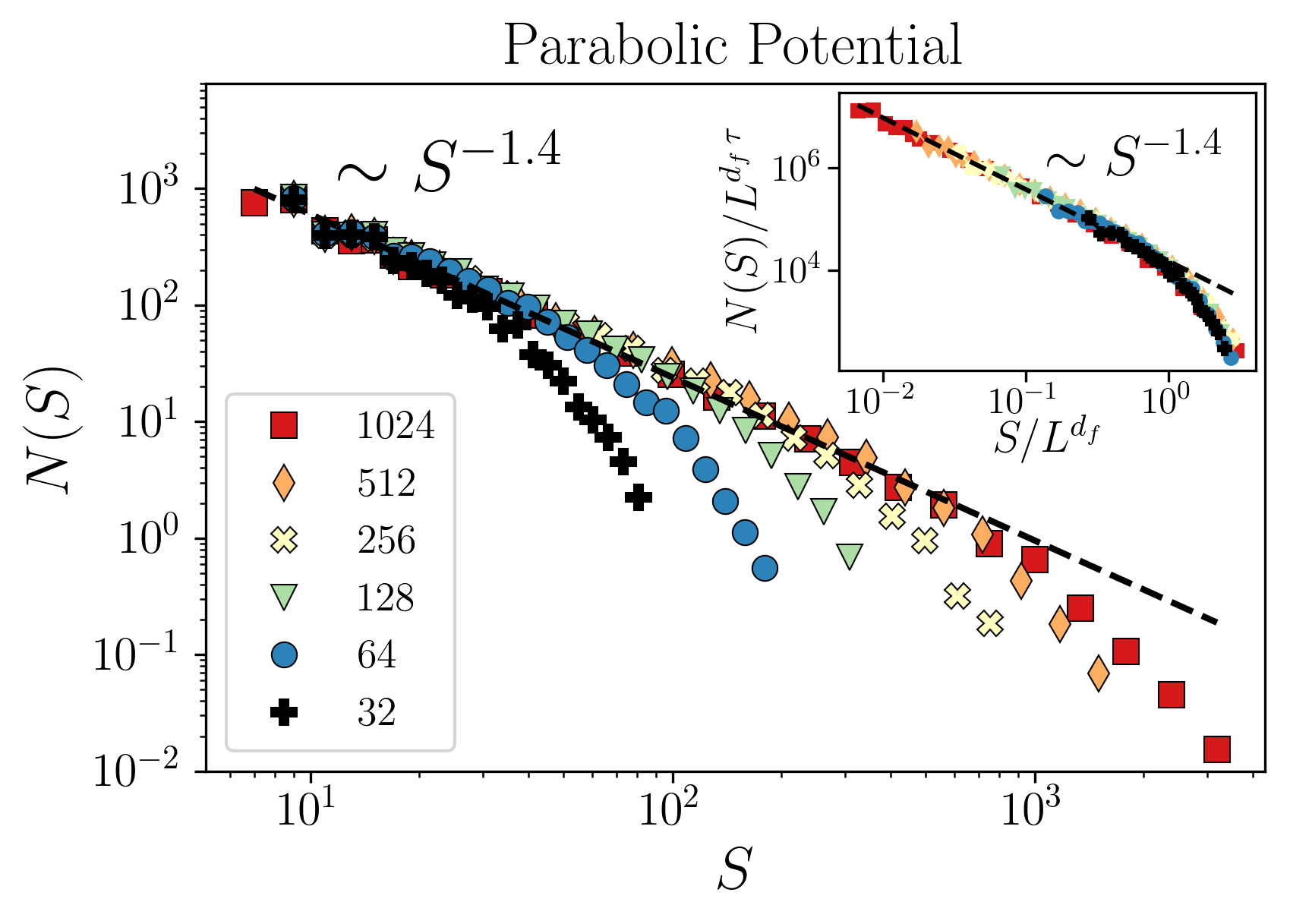}
		%\caption{Caption for image 1}   
%	\end{subfigure}   
	
%	\begin{subfigure}{.5\textwidth}  
%		\centering  
		\includegraphics[width=8cm,clip=true]{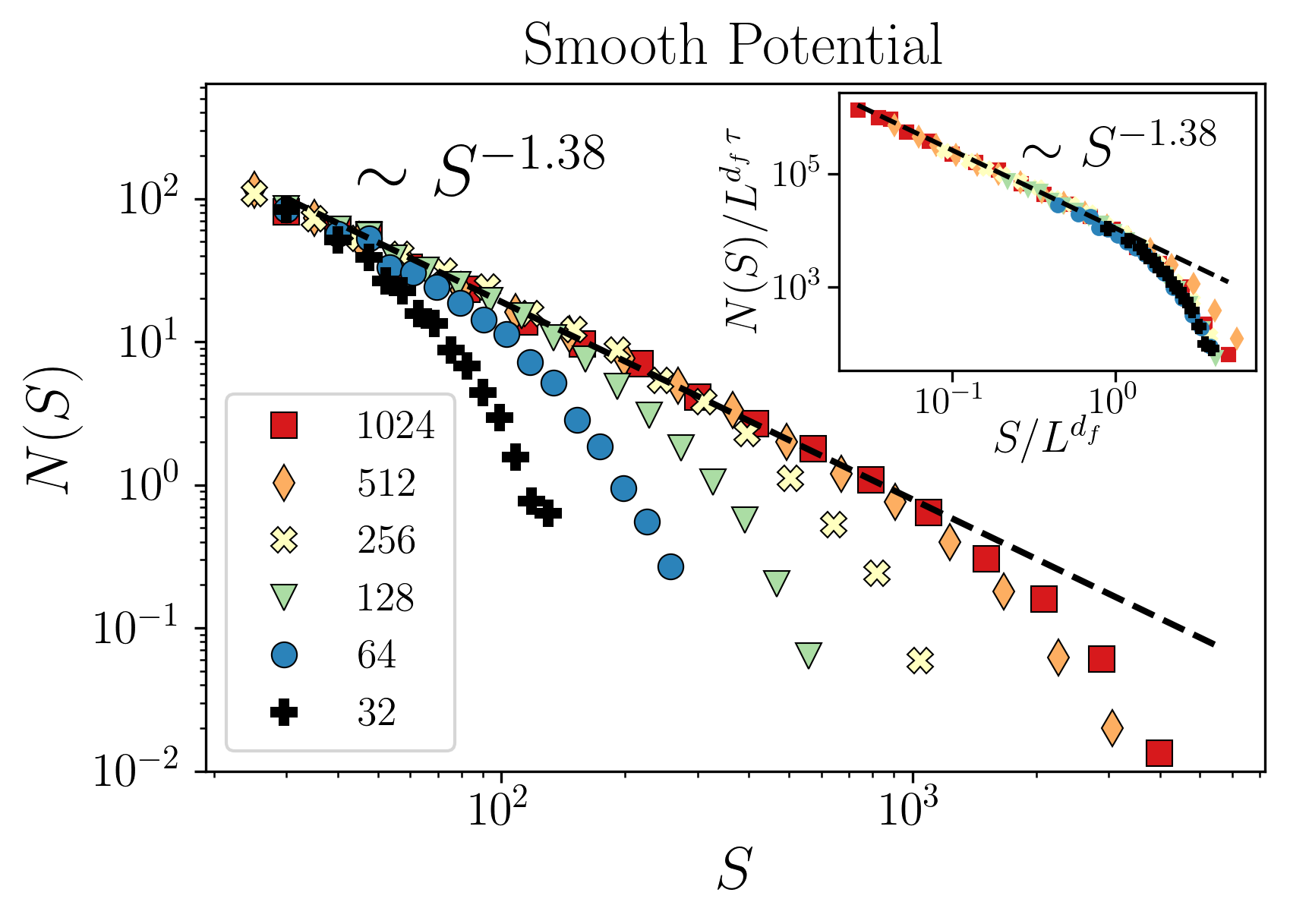}
		%\caption{Caption for image 2}   
%	\end{subfigure}  
	\caption{Histogram of avalanche size distribution, in systems of different sizes, for smooth and parabolic potentials. The dashed lines display the power law behaviour of the distributions. In the insets, the rescaling of avalanche size distributions using $d_f=1.01$ allows to collapse data from different system sizes.
	\label{Fig:tau}}  
\end{figure}

Results for the avalanche size distribution (Figure \ref{Fig:tau}) display a power law $P(S)\sim S^{-\tau}$, with $\tau \simeq 1.40$ for parabolic potentials and $\tau \simeq 1.38$ for smooth potentials. The power laws are cut off at large avalanche size by the system size. This cut off defines the fractal dimension $d_f$ of the avalanches which describes how the maximum size of avalanches grows with the linear size of the system as $S_{max} \sim L^{d_f}$. In order to determine $S_{max}$ most reliable from the simulation we use a relation of $S_{max}$ with the average size of $S$ and $S^2$ that reads\cite{Rosso_2009} $S_{max} \sim \overline{S^2}/(2 \overline{S})$.

We obtain from the simulations that ${d_f} \simeq 1.01$ for parabolic potentials and ${d_f} \simeq 1.01$ for smooth potentials. Taking into account the numerical uncertainties, we conclude that both $\tau$ and $d_f$ are independent on the potentials being of the smooth or parabolic type.

\begin{figure}[h!]  
%	\centering  
%	\begin{subfigure}{.5\textwidth}  
%		\centering  
		\includegraphics[width=8cm,clip=true]{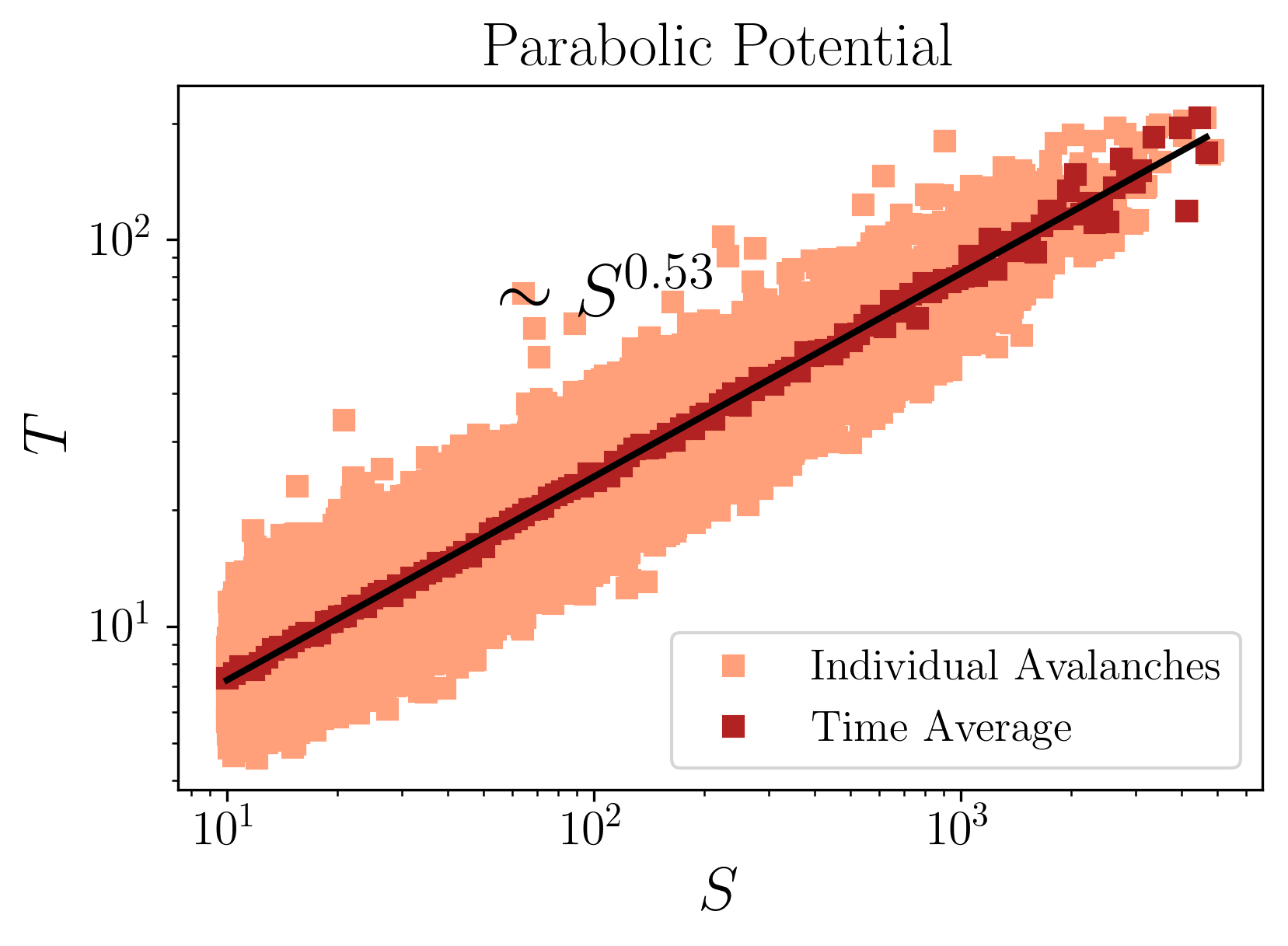}
		%\caption{Caption for image 1}   
%	\end{subfigure}   
	
%	\begin{subfigure}{.5\textwidth}  
%		\centering  
		\includegraphics[width=8cm,clip=true]{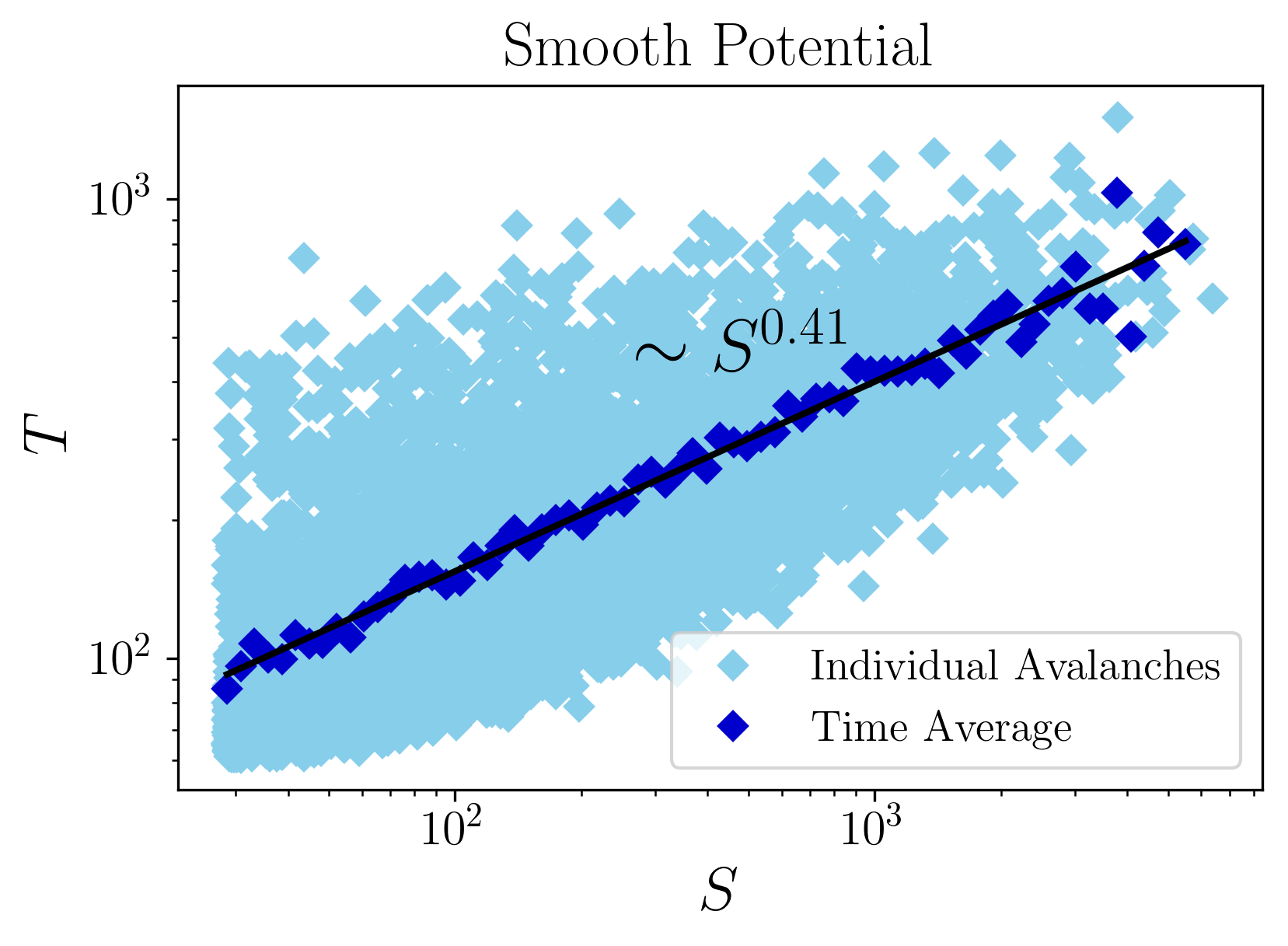}
		%\caption{Caption for image 2}   
%	\end{subfigure}  
	\caption{Avalanche duration vs. avalanche size, for both kinds of potential, in a system of $L=1024$. The light-color dots correspond to individual avalanches. The darkest dots correspond to an average over logarithmic width avalanche size windows, and they are shown in order to display the overall behaviour. Finally, black lines show the power law relation which on average is satisfied by the two quantities analyzed.
		\label{Fig:pes}}  
\end{figure}

In order to calculate the dynamical exponent $z$, we first plot the relation between duration and size of avalanches.
This is done in Figure \ref{Fig:pes}. The data show a wide dispersion, but averaging over avalanche size windows of logarithmic width, a well defined power law $T \sim S^p$ is obtained. The values of $p$ that are obtained from fitting are 
$p_p\simeq0.53$ for smooth potentials and $p_s \simeq 0.41$ for the parabolic potential. From this values and the definition of the dynamical exponent $z$ as  $z = p d_f$ \cite{Rosso_PNAS}, it is obtained that
 $z_p \simeq 0.53$ for smooth potentials and $z_s \simeq 0.42$ for parabolic potentials. We consider this difference to be significant, and our conclusion is that the dynamical exponent $z$ is different for parabolic and smooth potentials.

\begin{figure}[ht!]
	\includegraphics[width=7cm,clip=true]{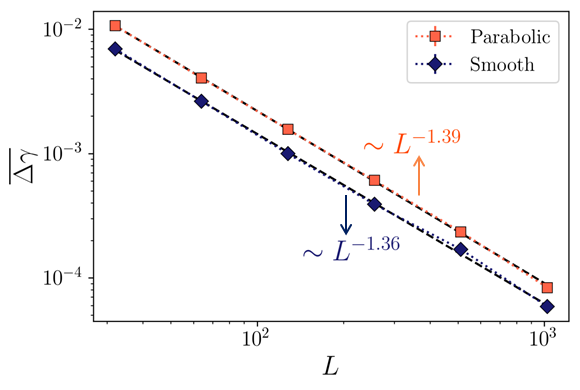}
	\caption{Average amplitude of strain increment $\Delta \gamma$ to trigger a new avalanche as a function of $L$. Black dashed lines are linear fits whose exponents allow to determine $\theta$ from $\overline {\Delta\gamma} \sim L^{\frac{-d}{1+\theta}}$.
	\label{Fig:Theta}}
\end{figure}

The last exponent that was calculated is the $\theta$ exponent, measuring the distribution of distance-to-instability at different position of the sample. If $x$ is the additional stress that has to be added to a given site to become unstable and jump to the next potential well, then $\theta$ is defined through the probability distribution of $x$ for $x$ close to zero, as $P(x) \sim x^{\theta}$. It is not straightforward (particularly in the smooth potential case) to calculate $\theta$ from a given equilibrium configuration in the system. However, the following trick can be used \cite{Rosso_PNAS}: $\theta$ can be calculated by following the average strain increase $\overline {\Delta\gamma}$ that has to be applied in order to activate consecutive avalanches. The results is that $\overline {\Delta\gamma} \sim L^{\frac{-d}{1+\theta}}$. By calculating $\overline {\Delta\gamma}$ for different values of $L$, $\theta$ can be determined. The corresponding graph is presented in Figure \ref{Fig:Theta} and we obtain that $\theta \simeq 0.44$ for parabolic potentials and $\theta \simeq 0.47$ for smooth potentials. The two values coincide within the numerical precision.

\begin{figure}[h!]  
%	\centering  
%	\begin{subfigure}{.5\textwidth}  
%		\centering  
		\includegraphics[width=8cm,clip=true]{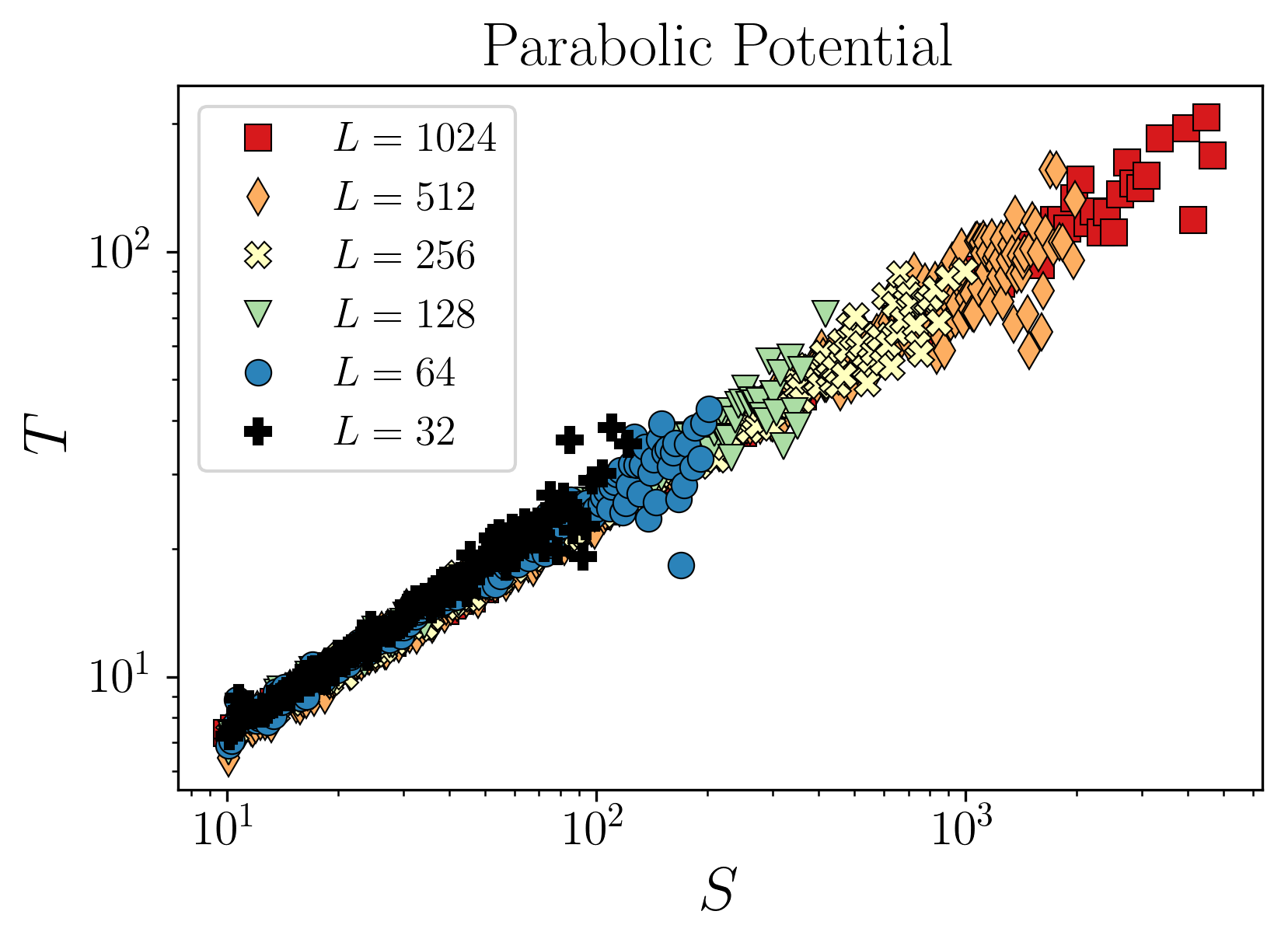}
		%\caption{Caption for image 1}   
%	\end{subfigure}   
	
%	\begin{subfigure}{.5\textwidth}  
%		\centering  
		\includegraphics[width=8cm,clip=true]{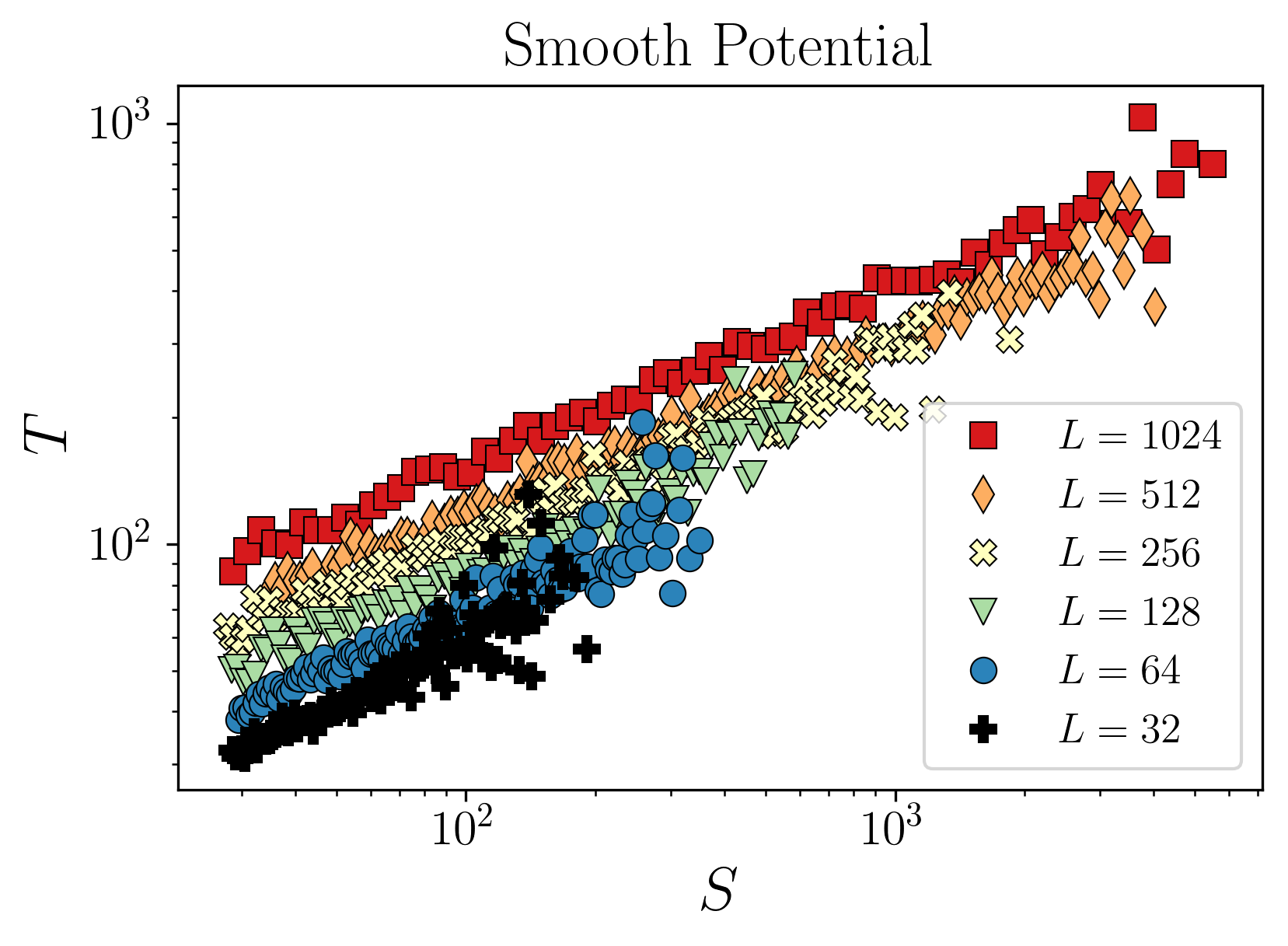}
		%\caption{Caption for image 2}   
%	\end{subfigure}  
	\caption{Average avalanche duration vs. avalanche size, in systems of different sizes, for both kinds of potential.
		\label{Fig:NUB}}  
\end{figure}

\begin{figure}[ht!]
	\includegraphics[width=8cm,clip=true]{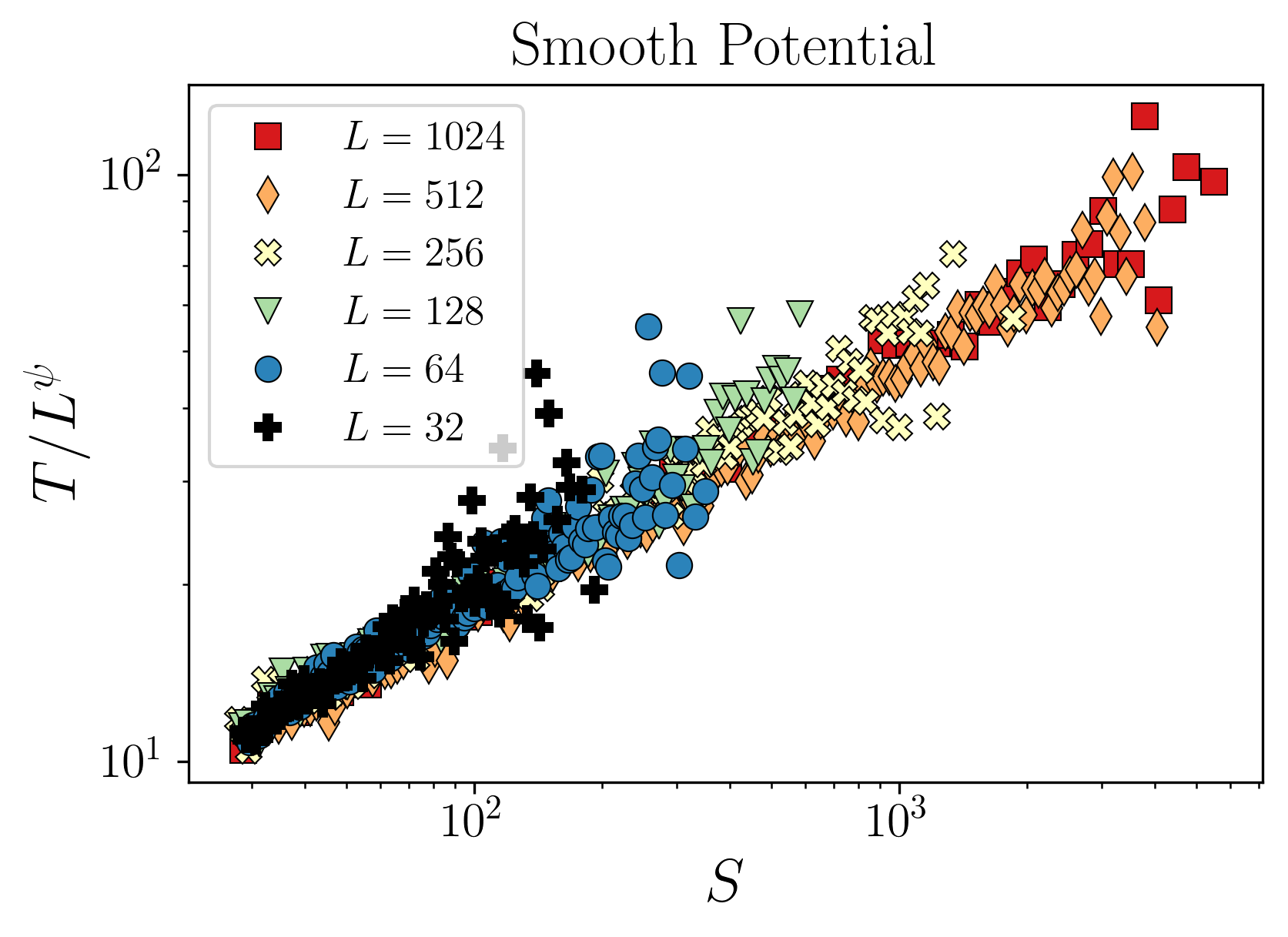}
	\caption{Normalized average avalanche duration vs. avalanche size, in systems of different sizes, for smooth potential. The shown curves correspond to $\psi=0.3$. The collapse of all these curves is an indication that the Eq. \ref{Tn} is well satisfied.
	\label{Fig:NUB_2}}
\end{figure}

The conclusion from the numerical simulations is that the ``dynamical" exponents $\beta$ and $z$ (those  that crucially depend on the time that particles take to jump between consecutive potential wells) depend on the kind of potential used, whereas static exponents such as $\tau$, $d_f$, and $\theta$ do not.\cite{footnote} The analysis of the next section rationalizes this behavior.
Yet, an additional unexpected difference between smooth and parabolic potentials was observed. Figure \ref{Fig:NUB} shows curves of average avalanche duration vs. size for different system sizes. In the parabolic case as the system size increases we simply observe that the data extend to larger values of $S$ and $T$. In the smooth potential case, we observe that the curves for different system sizes do not overlap
even for small avalanches. This indicates that there is a non-trivial dependence of the avalanche duration with $L$. If we suppose that the $L$ dependence can be factorized as a power of $L$, then we can define a normalized time as
\begin{equation}
T_{n}(S)=\frac{T(S,L)}{L^{\psi}}.
\label{Tn}
\end{equation} 
Using $\psi=0.30$ we obtain the curves for the normalized times seen in Figure \ref{Fig:NUB_2}. The collapse of all these curves is an indication that the Eq. \ref{Tn} is well satisfied.

The additional dependence of duration $T$ on system size $L$ that was observed in the case of smooth potentials allows an alternative definition of the dynamical exponent $z$. Instead of comparing the duration of avalanches with different sizes for a fixed $L$, we can compare the duration of the largest avalanche that occur for different system size. This allows to define an alternative exponent 
${z_s}^*=z_s+\psi$. Consequently, we have ${z_s}^* \simeq 0.72$.

\section{Mean field description}

We want to explore here the reasons why there are two different values of the dynamical exponents $\beta$ and $z$ for smooth and parabolic potentials, whereas those describing static properties, such as $\tau$, $d_f$, $\theta$ are the same.
Let us first analyze in more detail the form of the equations of the model (Eqs. \ref{modelo}-\ref{gdeq}) in real space. Since $G_{\bf q}\le 0$, we see that $G_{ii}\sim \sum_{\bf q} G_{\bf q}< 0$. Noting $k\equiv -G_{ii} $ we can write  Eq. (\ref{gdeq}) 
as
%generates a local term in real space. It is convenient to separate this term explicitly and write the real space form as
\begin{equation}
\dot e_{i}=-\frac{dV_i}{de_i}-ke_{i}+\sum_{j\ne i}G_{ij}e_{j}+\sigma
\label{tt2}
\end{equation}
Now we separate $G_{ij}$ in the last term as it average value, and its fluctuating part:
\begin{equation}
G_{ij}= \frac{k}{(N-1)} +\widetilde G_{ij}
\label{tt3}
\end{equation}
The equations of the model are then written as
\begin{equation}
\dot e_{i}=-\frac{dV_i}{de_i}+k(\overline {e_j}-e_{i})+\sum_{j\ne i}\widetilde G_{ij}e_{j}+\sigma
\label{t2}
\end{equation}

The kernel $\widetilde G_{ij}$ still has the $(r_i-r_j)^{-2}$ decay with distance, and the quadrupolar angular symmetry. But we emphasize that its spatial average vanishes: $\sum_{j\ne i}\widetilde G_{ij}=0$. %The average value of the original $G_{ij}$ (Eq. \ref{modelo}) has produced the appearance of the ``mean field like" term proportional to $k$ which couples strain at site $i$ to the mean value of the strain in the whole system.
Eq. \ref{t2} is appropriate to consider approximate treatments of the model.

%[In the Appendix we give an alternative derivation of a scalar model similar to the one presented here, in which we start from the beginning with a scalar description of the problem in the case in which the external driving is a single shear, instead of a deviatoric stress. The final model is similar to ... \textcolor{red}{[COMPLETAR]} except that the interaction kernel is not quadrupolar, but dipolar, reflecting the geometry of the imposed shear.]

\subsection{Naive mean field: the Prandtl-Tomlinson problem}

Neglecting the fluctuating term proportional to $\widetilde G_{ij}$ in Eq. (\ref{t2}), it reduces to 
\begin{equation}
\dot e=-\frac{dV}{de}+k(\overline {e}-e)+\sigma
\label{unapla}
\end{equation}
In order to obtain the flow curve ($\sigma$ vs $\dot \gamma$) in this limit, we note that $\overline e\equiv \dot\gamma t$. Defining also 
\begin{equation}
w(t) \equiv \dot\gamma t +\sigma/k
\label{wt}
\end{equation}
it is obtained:
\begin{equation}
\dot e=-\frac{dV}{de}+k(w(t)-e)
\label{unapla2}
\end{equation}
Written in this form, we see that $e$ is driven by the applied $w(t)$ on top of the potential $V(e)$ through a spring of constant $k$.
According to Eq. (\ref{wt}), the stress can be calculated as the average force on the driving spring: $\sigma=\overline{k(w(t)-e)}$.
This is just the Prandtl-Tomlinson (PT) model used to qualitatively describe the origin of a friction force between sliding solid bodies \cite{Prandtl,Tomlinson,Popov}. In the absence of thermal fluctuations --as it is the case here--, the PT model has a critical stress $\sigma_c$ for $\dot\gamma\to 0$ (as long as there are points at which $-d^2 {V(e_2)}/{d e_2^2}>k$ ), and a power law increase of $\sigma$ for finite $\dot\gamma$, i.e, $\dot\gamma\sim (\sigma-\sigma_c)  ^\beta$. The value of $\beta$ turns out to be dependent of the kind of potential that is used. For smooth potentials $\beta=3/2$, whereas for parabolic potentials (with points at which the first derivative has jumps) the value $\beta=1$ is obtained \cite{Narayan}. Namely, Eq. \ref{unapla2} provides a simple case in which the value of $\beta$ depends on the form of the potential. 

In order to qualitatively consider the avalanche statistics and its possible dependence on the kind of potential in this mean field approach, we will go back to Eq. (\ref{unapla}), and replace the uniform force $\sigma$ by driving at a constant speed $\dot \gamma$ through a spring of a small stiffness $k_0$. 

\begin{equation}
\dot e=f(e)+k(\overline e -e)+k_0(w(t)-e)
\label{1pla2}
\end{equation}
%\textcolor{red}{[agregué el "(t)" despues de w, en la ec de arriba]}
The time fluctuations of $\overline e$ are of order $1/N$ ($N$ being the total number of particles in the system). Although this fluctuation goes to zero in the thermodynamic limit (and then it reproduces the same average evolution than Eq. (\ref{unapla}) when $k_0\to 0$), it is enough to produce non trivial avalanches in the system. In fact, the statistics of the avalanches produced by a model like Eq. (\ref{1pla2}) are well known.
Avalanches distribute with a cut-off power law $P(S) \sim S^{-\tau} g(S/S_{max})$, where $g$ is a cut off function, $\tau=3/2$, and the cut-off value $S_{max}$ depends on $k_0$ as $S_{max}\sim k_0^{-2}$. As the value of $k_0$ is progressively reduced, avalanches with a critical size distribution $P(S)\sim S^{-3/2}$ are obtained. The 3/2 value of the $\tau$ exponent hold both for smooth and parabolic potentials. 

%A simple understanding of the origin of the $\tau=3/2$ result will be useful for the further understanding of avalanche durations. In the dynamical evolution according to Eq. (\ref{1pla2}), each site $i$ may be classified according to their distance to instability $\delta e_i$, representing the additional strain that has to be applied to site $i$ to make it jump to the 
%next potential well of $V(e)$. 
%An avalanche starts when external driving makes one site unstable. By jumping to the next potential well, this site
%increases its $e$ value a quantity of order 1, and produces an increase of $\overline e$ in an amount $\delta e\sim 1/N$. This produces a reduction of all $\delta e_i$ in the same amount. If now some sites become unstable, 
%the process is continued until there are no more unstable sites. 
%This is the epitome of a Galton-Watson branching process, in which an active site produces additional activations with some fixed probability. The number of active sites as a function of the number of sites that have already passed to the next potential well is thus simply a random walk process. When the number of active sites becomes zero the avalanche stops. The statistics of avalanche sizes is then the statistics of zero crossings of a random walk, and this provides $P(S)\sim S^{-\tau}$, with $\tau=3/2$.

However, differences appear between smooth and parabolic potentials when considering the duration of the avalanches.
For the calculation of this time, it becomes crucial to take into account the time that an unstable site actually takes  to move to the new equilibrium position in the next potential well. In the case of parabolic potentials, the pushing force is finite as soon as the instability point is overpassed, and this implies that this time is independent of the stress excess $\Delta \sigma$. 
The situation is different for smooth potentials. A site that becomes unstable feels a pushing force that is 
strongly dependent on the stress excess $\Delta \sigma$, over the threshold for instability. It turns out that the time an unstable site takes to reach the new equilibrium position at the next potential well scales as $\sim \Delta \sigma ^{-1/2}$ \cite{Strogatz}.

An analysis based on this difference between smooth and parabolic potentials (to be presented elsewhere \cite{arj}) leads to the conclusion that
for parabolic potentials, avalanche duration $T$ scales with the avalanche size $S$ as
$T\sim S^{1/2}$, whereas for smooth potentials $T\sim S^{1/4}$. 

Thus in addition to $\beta$, the dynamical exponent $z$ is different for parabolic and smooth potentials in mean field.
This is a remarkable result. It shows that even in mean field, and in addition to the already discussed difference in the $\beta$ exponent, there are differences in the dynamical exponent when comparing parabolic and smooth potentials. We remark that the static exponents %($\nu$ and $\eta$, or $d_f$) 
are the same for both potentials in mean field. The exponents that are different are those
related to the dynamical characteristics of the avalanches, and the difference originates in the qualitatively different way in which a particle jumps from one potential well to the next, for smooth or parabolic potentials.

\subsection{A ``dynamical" mean field approach: stochastically driven Prandtl-Tomlinson particles}

By introducing the definition of $w(t)$ into Eq. (\ref{t2}) we obtain
\begin{equation}
\dot e_{i}=-\frac{dV_i}{de_i}+k(w(t)-e_{i})+\sum_{j\ne i}\widetilde G_{ij}e_{j}
\label{qsy}
\end{equation}
This defines a set of coupled PT models, in which the variable $e_i$ evolves under the external uniform driving $w(t)$ on the potential $V_i$, and it is affected by all other $e_j$ through the coupling term $\widetilde G_{ij}$. 
We will now make a description in which this term is decoupled and treated as an external perturbation. 

To begin with, we start with a brief digression.
The accuracy of a mean field approximation depends essentially on the range of the interaction. Let us consider for the moment a standard, ferromagnetic Ising model in two spatial dimensions, with interactions decaying as $1/r^\alpha$. The values of the critical exponents depend continuously on the value of $\alpha$ and move towards mean field values as $\alpha$ is reduced. When $\alpha=2$ (in general, when $\alpha$ is equal to space dimensionality), the model becomes mean field and the critical exponents are exactly given by their mean field values. 
A simple way to understand this result is the following. A spin in a given position interacts with a weighted sum of all other spins in the system. For $\alpha> 2$ the influence of any individual spin on this sum has a finite (non-zero) weight. However for $\alpha\le 2$, the influence of any individual spin on the effective field seen by any other spin is infinitesimal (for an infinite size system). This means that fluctuation effects are unimportant, and mean field results are exact. Note that this implies not only that the exponents are mean field, but that the full solution to the problem is exactly given by the mean field approximation.

In our case, the interaction term in Eq. \ref{qsy} has precisely the $\sim 1/r^2$ decay. However the sign is alternating with zero average, and this prevents the application of the arguments of the previous paragraph in a direct form. On average, the mean value of the last term is zero, it is its fluctuation in time what is relevant. In this sense, we note that although the sum of the last term is zero on average, the contribution of any individual $e_i$ to the {\em fluctuation} is still infinitesimal when $\alpha\le 2$. This suggests that a ``mean field" description should be rather accurate, if not exact, in the present case too. In this context the meaning of ``mean field" is that the last term can be treated as an externally given fluctuating term, and in this way the evolution of each local variable becomes a one particle problem.

In other words, we will write formally Eq. \ref{qsy} as
\begin{equation}
\dot e_{i}=f_i(e_{i})+k(w(t) -e_{i})+\xi_i(t)
\label{tomlinson3}
\end{equation}
where 
\begin{equation}
\xi_i(t)=\sum_{j\ne i}\widetilde G_{ij}e_{j}
\label{xi}
\end{equation}
Now, $\xi_i(t)$ will be taken to be an external noise. 
In the end, we should require this noise to be compatible with the evolution of the local variables, i.e., Eq. (\ref{xi}) be satisfied. However, as a first step we will consider Eq. (\ref{tomlinson3}) on its own, assuming some statistical properties of the stochastic noise $\xi_i(t)$.

Taking into account that according to Eq. (\ref{xi}) the time evolution of $\xi_i$ depends on the variation rate of $e_j$, the statistical properties of the noise term must scale with the velocity at which the system is driven. 
We are interested mainly in the case in which driving is very slow.
In this limit, $\xi_i$ can be considered to depend directly on the control variable in the system, that is, on the applied external strain $\dot \gamma t$.
This means that the dependence of $\xi_i$ on the strain rate can be explicitly incorporated by writing:
\begin{equation}
\dot e_{i}=f_i(e_{i})+k(w t -e_{i})+\xi_i(\dot \gamma t)
\label{t3}
\end{equation}
This equation defines what we call the stochastically driven Prandtl-Tomlinson model. The evolution of $e_{i}$ will depend on the amplitude and correlations of the noise term $\xi$, as well as on the form of the force $f_i(e_{i})$.

We will consider the case (that will be shown is relevant in the yielding context) of a $\xi(x)$ noise that has self-similar correlation properties characterized by the so called Hurst exponent $H$. This means that, statistically
\begin{equation}
\xi_i(\lambda x) \sim \lambda^{H} \xi_i(x)
\label{th}
\end{equation}
%\textcolor{red}{[Aca por ser el cumulativo es H+1. LA DEFINICION DEL H DE UNA SEÑAL NO LLEVA EL +1. NO IMPORTA SI ES ACUMULATIVO O NO. SI NO NO TE DA BIEN PARA EL RW COMUN]}
Note that a standard random walk has $H=1/2$.

%The Hurst exponent is the analog for time signals of the roughness exponent of interfaces, or other spatial signals(?).
%The value of $H$ goes between $0<H<1$. 
%When $H=1/2$, 
%$\eta(x)$ is a standard random walk....
%Values of $H$ higher (lower) 
%than $1/2$ correspond to signals with a positively (negatively) correlated memory.

The flow exponent of the model defined by Eqs. (\ref{t3}) and (\ref{th}) was worked out in
Ref. \cite{Jagla_P-T}. There it was shown that 
\begin{equation}
\dot{\gamma} \sim (\sigma-\sigma_c)^\beta,
\label{Ec:beta_en_ja}
\end{equation}
with a flow exponent
\begin{equation}
\beta=\frac{1}{H}-\frac{1}{\alpha}+1
\label{Ec:beta_en_ja_2}
\end{equation}

where $\alpha$ is related to the analytic form of the potential at the transition point between consecutive potential wells: $\alpha=1$ for parabolic potentials and $\alpha=2$ for smooth potentials. Although we do not know for the moment what the appropriate value of $H$ is, we note that from Eq. (\ref{Ec:beta_en_ja_2}), the difference between $\beta$ values for smooth and parabolic potentials is 1/2, independently of the value of $H$. This is well satisfied by the results of the full simulations presented in Section III.

The present independent particle analysis gives a prediction also on the value of the $\theta$ exponent in the system.
We remind that this exponent characterizes the equilibrium distribution of distances $x$ to the instability point. This distribution $P(x)$ is expected to behave as $P(x)\sim x^\theta$ for small $x$. If there is no stochastic term in the driving ($\xi=0$ in (\ref{t3})) the value of $x$ reduces linearly in time until destabilization, the distribution $P(x)$ is flat and we obtain $\theta=0$. If there is a stochastic term in the driving the value of $\theta$ is determined from the distribution $P(x)$ of a Fractional Brownian Motion with an absorbing wall at $x=0$, which is \cite{majumdar1,majumdar2} $P(x)\sim x^{\frac 1H-1}$, i.e., 
\begin{equation}
\theta=\frac 1H-1. 
\end{equation}
Note that this result is independent of the potential being of the parabolic or smooth type.
It is interesting to eliminate the (still undetermined) value of  $H$ from the expressions of $\beta$ and $\theta$ to obtain that in this mean field situation they are related by 
\begin{equation}
\beta=\theta+2-\frac 1 \alpha
\label{betatheta}
\end{equation}
We notice again that the numerical values obtained in Section III for $\theta$ and $\beta$ quite closely satisfy Eq. (\ref{betatheta}).

\subsection{Finding the value of $H$}

\begin{figure}[ht!]
	\includegraphics[width=8cm,clip=true]{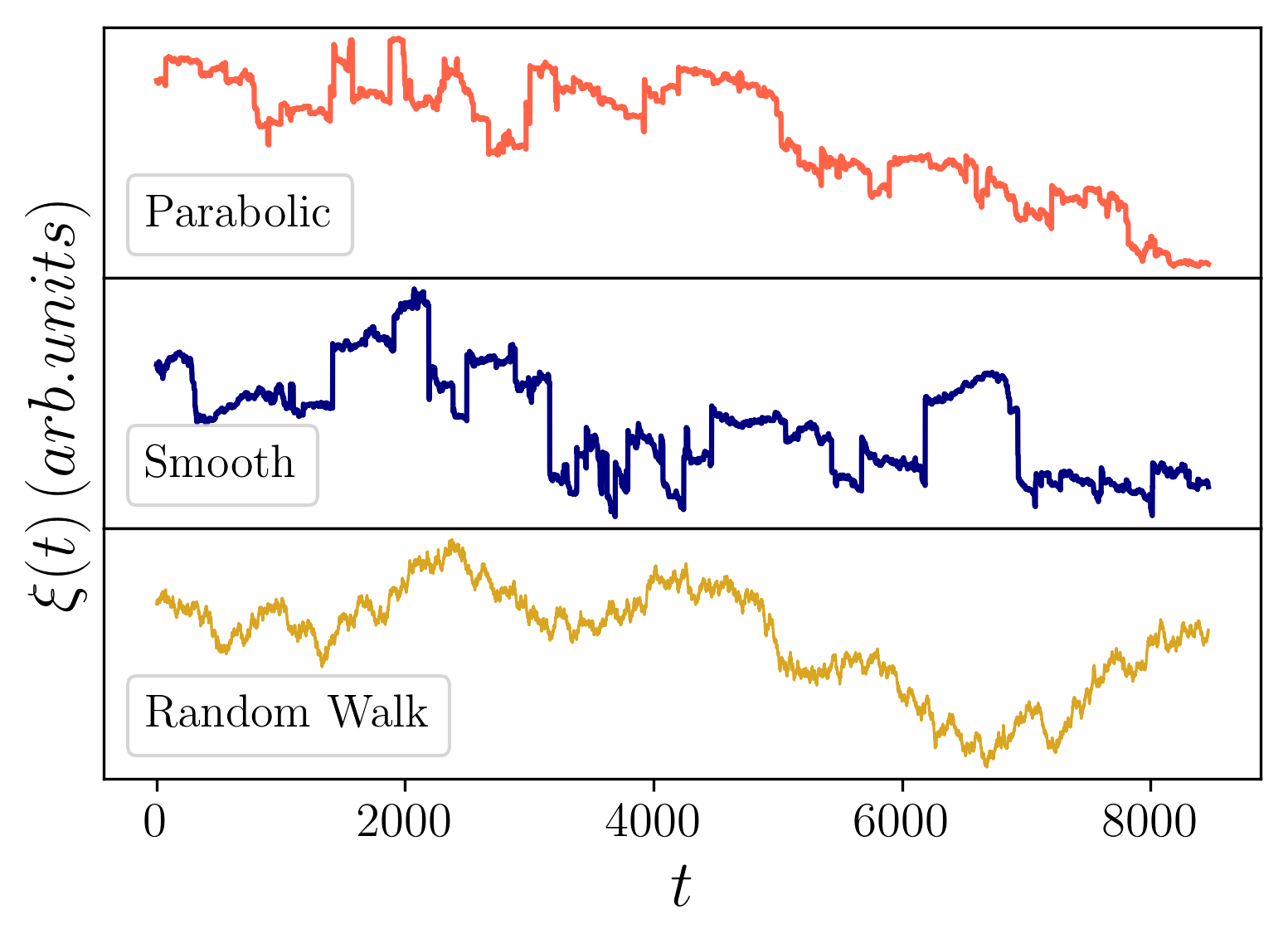}
	\caption{Typical functions $\xi(t)$ evaluated to Eq. \ref{xi} for the parabolic and smooth case, in a quasistatic simulation of the model described in section III. The time variable represents the number of avalanches. For comparison a standard random walk is also displayed.
	\label{Fig:DFA0} }
\end{figure}

To check the consistency of our approach, and in particular the value of $\beta$ predicted by Eq. (\ref{Ec:beta_en_ja_2}) we must calculate the $H$ value of the signal $\xi$. In order to do this, we generate time series of $\xi$ according to its definition as given by Eq. (\ref{xi}) running a full simulation of the model as described in Section III.
The simulation is done in a quasistatic case, with $\dot\gamma\to 0$. Examples of the $\xi(t)$ signals that are obtained both in the parabolic and smooth cases are shown in Fig. \ref{Fig:DFA0}. 
%\textcolor{red}{Finally, we calculate the Hurst exponent of different signals
%%	, which include the ones shown in Fig. \ref{Fig:DFA0}, 
%using the Detrended Fluctuation Analysis technique \cite{DFA_1,DFA_2,DFA_3}.  
%Essentially, the method studies how the signal growths within windows of different widths allowing to determine the value of $H$. In order to improve our results, we do this analysis over five signals that correspond to a five equi-spaced points of the system for each potential. 
%The average result obtained (Figure \ref{Fig:DFA} shows one of the results)
%is $H \simeq 0.67$ for parabolic potentials and $H \simeq 0.64$ for smooth potentials. Within the numerical errors the two values coincide. This should come with no surprise at this point, since we do not expect differences between the two kind of potentials in the quasistatic limit. Moreover, when plugged into Eq. (\ref{Ec:beta_en_ja_2}) this value of $H$ provides $\beta$ values for parabolic and smooth potential that perfectly fit those obtained in the numerical simulations of Section III.}
Overall, we generate five signals that correspond to points of different regions of the system, for each potential. The Hurst exponent of the signals is then obtained using the Detrended Fluctuation Analysis technique \cite{DFA_1,DFA_2,DFA_3}. Essentially, the method studies how the signal growths within windows of different widths allowing to determine the value of $H$. The average result obtained is $H \simeq 0.67$ for parabolic potentials and $H \simeq 0.64$ for smooth potentials. Within the numerical errors, the two values coincide. This should come with no surprise at this point since we do not expect differences between the two kinds of potentials in the quasistatic limit. Moreover, when plugged into Eq. (\ref{Ec:beta_en_ja_2}) this value of $H$ provides $\beta$ values for parabolic and smooth potential that perfectly fit those obtained in the numerical simulations of Section III.

%\begin{figure}[ht!]
	%\includegraphics[width=8cm,clip=true]{DFA.png}
	%\caption{\textcolor{red}{Results from the Detrended Fluctuation Analysis %technique over one of the five signals obtained for each potential.}
	%%%%, which are similar to those shown in the Fig. \ref{Fig:DFA0}}.
	%The slope determines the value of $1+H$. \textcolor{red}{The average result %of the values that were acquired by linear fitting gives $H \simeq 0.67$ %for parabolic potentials and $H \simeq 0.64$ for smooth potentials}. For %the standard random walk the classical value $H=0.5$ is obtained.
%	\label{Fig:DFA} }
%\end{figure}

It remains to be understood why a value $H\simeq 0.65$ shows up in the simulations.
In oder to address this point, we notice that according to its definition in Eq. (\ref{xi}), the fluctuating term $\xi(t)$ gets a cumulative contribution every time an avalanche occurs in the system. So we can try to make an estimation of the form of $\xi(t)$ by assuming a random and uncorrelated distribution of avalanches in the system, with a size distribution characterized by some exponent $\tau$.
Each avalanche will generate a contribution to $\xi$ that we note $\delta \xi$.
Under the assumption of uncorrelated avalanches, we can determine the distribution on increments $P(\delta \xi)$. If $P(\delta \xi)$ happens to have long tail, namely 
\begin{equation}
P(\delta \xi) \sim \frac{1}{|\delta \xi| ^{\nu+1}}
\label{Dist}
\end{equation}
(with $\nu<2$) for large $|\delta \xi|$, then its random accumulation will produce a generalized random walk $\xi(t)$ characterized by a non trivial Hurst exponent 
where $H=1/\nu$ \cite{Lin_Wyart}. %\textcolor{red}{[No se si es la correcta referencia]}

\begin{figure}[ht!]
	\includegraphics[width=6cm,clip=true]{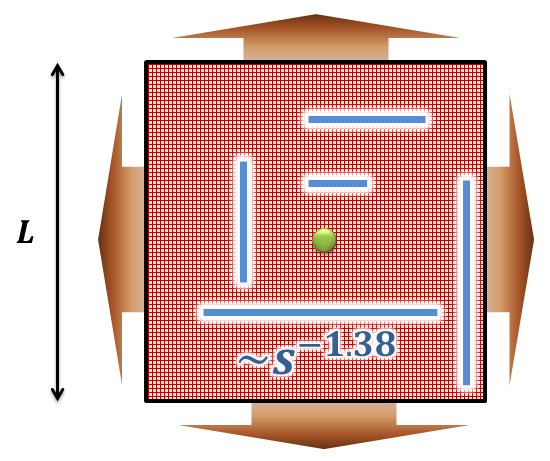}
	\caption{Effect of avalanches on the stress of a given site. The stress increment on the central site (green dot) produced by avalanches triggered everywhere in the system (blue lines) is calculated 
taking into account the Eshelby interaction. Avalanches are assumed to be linear objects along the two orthogonal easy directions in the system, and are distributed with the known power law size distribution, and uncorrelated in time and space. The brown arrows indicate the periodic boundary conditions used.
	\label{Fig:WY} }
\end{figure}

\begin{figure}[ht!]
	\includegraphics[width=8cm,clip=true]{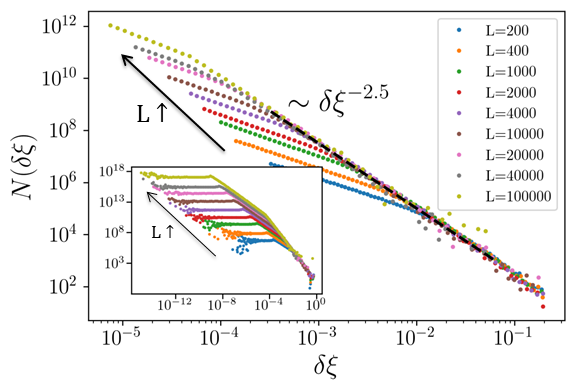}
	\caption{Histogram of stress increments produced by avalanches in a site of the system, in systems of different size.  The curves are vertically shifted in order to improve the visualization of results. The main plot shows the relevant region of the histograms which allows obtaining the Hurst exponent. The dashed line shows a reference slope and it corresponds to a $H=2/3$. The inset presents the full histograms.
	\label{Fig:wy_dis} }
\end{figure}

We consider a square system of linear size $L$ with periodic boundary conditions (Figure \ref{Fig:WY}), and we focus on the effect of random avalanches in the system on the strain at the central point. Avalanches are assumed to occur with a size distribution $S^{-\tau}$ and in the two orthogonal easy directions in the system. The goal is to calculate the strain increment on the central site produced by each avalanche and mediated by the Eshelby propagator. 
In principle this problem reduces formally to the calculation of a (three dimensional) integral, but we have not been able to find a closed form of the result, so we first show the result obtained using a Monte Carlo method.
The numerical implementation of this process generates the form of $P(\delta \xi)$ observed in Fig. \ref{Fig:wy_dis}.
$P(\delta\xi)$ displays a power law for large $\delta\xi$ that becomes more robust as the system size is increased. The value of the decay exponent depends slightly on the value of $\tau$ from which the avalanches were chosen, but using the actual value of $\tau$ ($\tau\sim 1.4$), we find $P(\delta\xi)\sim \delta\xi^{-(\nu + 1)}$ with $\nu\simeq 1.5$ and thus $ H\simeq 0.65$,
which coincides with the value directly determined from the numerical simulations through (Eq. \ref{xi}).

In Ref. \cite{Lin_Wyart}, Lin and Wyart also considered the strain fluctuations at a given site caused by the rest of the system (for the case of elasto-plastic models, and thus akin to our case of parabolic potentials, see below). Then, in a mean field approach, they 
were able to link the exponent $\nu$ in the distribution $P(\delta\xi)$ with the flow exponent $\beta$, finding a relation that is compatible with our results for parabolic potentials, namely $\beta=\nu$ for $1<\nu<2$ (if $\nu=1$ they find the flow curve has logarithmic corrections).
However, they calculate the statistics of $\delta\xi$ assuming it is formed by random kicks from individual sites, with an intensity given by the Eshelby kernel, $\sim\pm 1/r^2$, finding $P(\delta\xi)\sim \delta\xi^{-2}$, i.e, $\nu=1$, which is not the result we obtain. 
The reason of the difference is  that the contributions $\delta\xi$ that generate the stochastic noise cannot be considered as generated in isolated points, since they are typically produced by avalanches, that are extended objects. 

The effect of avalanche size distribution on the value of $\nu$ can also be estimated using the following argument.
An avalanche of size $S$ located at a distance $D$ from the origin produces an increase $\delta\xi$ of strain at the origin with the following characteristics. If $D \gg S$ then $\delta\xi\sim \pm S/D^2$ (in this case the avalanche behaves as a point-like object, the $\pm$ sign is a short hand for the angular dependence of the Eshelby interaction). This behavior prevails until $ D\sim S$. However for $D\lesssim S$ the effect of the avalanche 
becomes proportional to  $1/S$, and independent of $D$, as a direct integration shows. 
Then the contribution to $P(\delta \xi)$ from avalanches with a fixed value $S_0$ takes the form:
\begin{eqnarray}
P(\delta \xi|S_0)\sim \frac{S_0}{|\delta\xi|^2}~~~~\mbox {for}~ \delta\xi\le\frac 1S_0\\
P(\delta \xi|S_0)\sim S_0^2{\delta_\mathsf{D}}(\delta \xi-1/S_0)~~~~\mbox {for}~ \delta\xi\simeq \frac 1S_0
\end{eqnarray}
where $\delta_\mathsf{D}$ notes a Dirac delta function.
Now the total $P(\delta \xi)$ is obtained integrating this result over $S_0$, considering the probability distribution of $S_0$:
\begin{equation}
P(\delta \xi)=\int  P(\delta \xi|S_0)S_0^{-\tau} dS_0 
\end{equation}
The result is 
\begin{equation}
P(\delta \xi)\sim \frac 1{|\delta \xi|^{4-\tau}}
\end{equation}
which modifies the result obtained in \cite{Lin_Wyart} in the right direction: The value of $H$ is $H=1/(3-\tau)$ which provides (using $\tau\sim 1.4$ from the simulations in Section III) $H\simeq 0.62$, quite close to the value $H\simeq 0.65$ directly measured before.

All these verifications of self-consistency indicate that the treatment of the interaction term in Eq. (\ref{t2}) as a mean field fluctuating noise is a consistent and quantitatively accurate approach.

\section{Relation with elasto-plastic models}

It was already mentioned that one of the characteristics of the present model is that there is a single strain variable $e$, and no clear cut separation is made between elastic and plastic strains, contrary to what is usually done in EP models. However, for the case of parabolic potentials
this separation can in fact be proposed, and it is possible to discuss in detail the relation between EP models and the present one.

We consider our model with parabolic potentials. In this case, the central position of parabola at site $i$ (to be noted $\gamma^{pl}_i$) can be identified with the plastic deformation at site $i$, and Eq. (\ref{modelo}) can be written as

\begin{equation}
\dot e_{i}=\mu(\gamma^{pl}_i-e_i)+\sum_j G_{ij}e_{j}+\sigma
\label{hlt}
\end{equation}
where $\mu$ is the curvature of the potential. Note that this curvature is assumed to be equal at every potential well.
We will suppose that $\dot\gamma$ is so small that it can always be assumed that $e_i$ is in an equilibrium position. i.e., $\dot e_i=0$.
In this case we can write 
\begin{equation}
\sigma_i\equiv-\mu(\gamma^{pl}_i-e_i)=\sum_j G_{ij}e_{j}+\sigma
\label{si}
\end{equation}
where the local stress $\sigma_i$ has been introduced. Since all parabola have the same curvature, if the average strain increases at a rate $\dot\gamma$, the value of $\sigma_i$ increases uniformly in the system with the same rate, as long as no particle goes out of its local parabola. Namely
\begin{equation}
\delta \sigma_i=\mu\dot\gamma \delta t
\label{lineal}
\end{equation}
If $\sigma_i$ becomes larger than the maximum stress that site $i$ can sustain, the corresponding $\gamma^{pl}_i$ changes to  a new value $\gamma^{pl}_i+\delta \gamma^{pl}_i$ and the strains $e_i$ will accommodate to new values satisfying Eq. (\ref{si}). Upon changes in $\gamma^{pl}_i$, the corresponding changes $\delta \sigma_i$ in the stresses can be obtained from that equation. Working in Fourier space the result is
\begin{equation}
\delta \sigma_{\bf q}=\frac{\mu G_{\bf q}}{\mu-G_{\bf q}} \delta e_{\bf q}\equiv H_{\bf q}e_{\bf q}
\label{sigmagamma}
\end{equation}
where $G_{\bf q}$ is given in Eq. (\ref{gdeq}). Since the denominator is strictly positive for all ${\bf q}$, $H_{\bf q}$ still has the same zero modes that the original $G_{\bf q}$ and its $\cos(4\theta)$ symmetry, and being independent of the norm of ${\bf q}$ (as $G_{\bf q}$ itself), it has a decay in  real space as $1/r^2$.

In this way, the previous equation gives the effect of an increase in plastic deformation on the stress in the sample. The kernel for this influence has  the  Eshelby structure $\sim \cos(4\theta )/r^{2}$. Such an influence of the plastic strain on the stress (Eq. \ref{sigmagamma}), plus the linear increase of stress with applied strain (Eq. \ref{lineal}), are exactly the ingredients used for instance in the implementation of EP models given by \cite{Rosso_PNAS}. 
Yet an additional consideration is necessary. In EP models it is typically (sometimes implicitly) assumed that there is a fixed time scale for a site that has overpassed its maximum stress, to move to a state with $\sigma\simeq 0$ (this typically occurs at a constant rate, or in a single time step).
In our case, plastic strains change instantaneously when strain reaches the crossing between successive parabola, however, the starting Eq. (\ref{hlt}) has in fact a typical time scale $\tau \sim 1/\mu$ for an unstable site to reach its new equilibrium position. It is thus clear that qualitatively, our model with parabolic potentials can be interpreted as an elastoplastic model, and then it is not surprising that we get the same critical exponents as found for instance in \cite{Rosso_PNAS}.
The present comparison also suggests that the phenomenology of our model with smooth potentials might not be captured by usual elastoplastic models. But at the same time it suggests the appropriate modification in EP models to match this case too. In fact, the main difference between the dynamics of smooth and parabolic potentials seems to be the different time that it takes for a particle at a given potential well to reach the next one when it jumps over the barrier. For parabolic potentials, as we argued before, this time $\tau$ is roughly constant, independently of the stress excess over the critical value. This is what leads to consider a constant transition rate, and what makes possible the comparison with standard EP models. For smooth potentials however, the time $\tau$ that it takes to reach the new equilibrium position strongly depends on the stress excess $\sigma_i-\sigma_i^c$ over the critical value $\sigma_i^c$, actually $\tau \sim (\sigma_i-\sigma_i^c)^{-1/2}$\cite{Strogatz}. In an implementation in terms of transition rates, smooth potential would require to consider stress dependent transition rates. We think that with this additional ingredient EP models can be used to reproduce also the results we obtain here with smooth potentials.

Finally, we note that there are other kinds of EP models (such as the model of Picard \cite{picard}, or that used by Barrat {\em et al.} \cite{barrat}) that are directly defined in terms of its dynamics and cannot be derived from the minimization of a Hamiltonian function. It is possible, but not proved at present, that a modification of the transition rate in these models would also produce a change in the dynamical exponents $\beta$ and $z$ as we observed in our case changing from parabolic to smooth potentials.

\subsection{Relation to the H\'ebraud-Lequeux mean field}

The H\'ebraud-Lequeux model \cite{H-L}  is a further simplification on an elasto-plastic model, in which any plastic rearrangement is assumed to 
produce a random variation of stress on any other site. Note that the value of the random variation is renewed if the same site yields plastically a second time.

In our model as described by Eq. (\ref{qsy}) this random effect can be mimicked by replacing
in the last term the kernel $\widetilde G$ by a random coupling that is renewed every time $e_i$ jumps to a new potential well. It is clear that this produces a noise term $\xi(\dot\gamma t)$ as in Eq. (\ref{tomlinson3}) that is the accumulation of random contributions from all the strain jumps that occurred all across the system, i.e., a standard random walk, with a Hurst exponent $H=1/2$. According to our previous analysis, we know that this case provides (for parabolic potentials) $\beta=2$, $\theta=1$ in fact, similar to the values that are obtained in the H\'ebraud-Lequeux model.
Note that this approximation for the case of smooth potentials produces a value $\beta=5/2$ instead.

\section{Conclusions}

In this paper we have investigated the critical properties of the athermal yielding transition in a two-dimensional model that includes structural disorder and long range elastic interactions as two main ingredients. Our results strongly suggest that some critical exponents depend on the form of the plastic disorder potential, finding differences between the cases of a ``smooth'' potential case (in which minima are smoothly connected) and a ``parabolic'' case (in which the potential ia a concatenation of parabolic pieces) with discontinuous forces at the transition points. The exponents that differ between the two cases are the flow exponent $\beta$ and the dynamical exponent $z$. Other exponents are the same in the two cases. We interpret the differences as a consequence of the qualitatively different dynamics of the system around the transition points in the smooth and parabolic cases. We also claim that contrary to what happens in the depinning problem (where the two kind of potentials are known to produce no difference in the critical properties)
here the difference remains because of the long range nature of the elastic interaction. In fact, this long range nature of the interaction transforms the problem into an effective mean field one. We constructed explicitly the mean field theory describing the problem and showed it corresponds to a particle driven on top of the disordered plastic potential. The driving incorporates the mechanical noise of all other sites in the system as a stochastic contribution. We gave the values of most of the critical exponents in terms of the statistical properties of this noise, particularly its Hurst exponent $H$. As a consistency check we measured directly the value of $H$ in the full simulation and also estimated it from a simplified analysis, finding $H\simeq 2/3$. Overall, the values of the critical exponents found both for parabolic and smooth potentials, and the value of $H$ are totally consistent, giving support to our mean field interpretation of the transition.

\section{Acknowledgments}

We thank Ezequiel Ferrero, Jean-Louis Barrat, Vivien Lecomte and Alejandro Kolton for helpful discussions.

\appendix

\section{An alternative derivation of a scalar model under the assumption of single shear}

\begin{figure}[ht!]
	\includegraphics[width=6cm,clip=true]{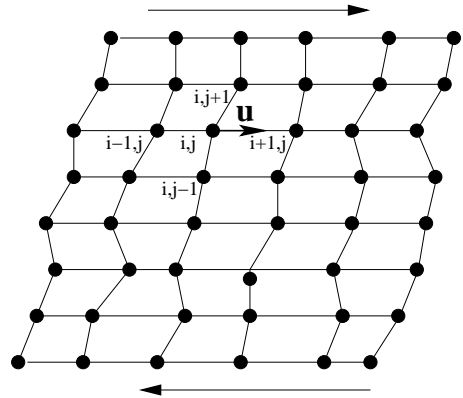}
	\caption{Geometry assumed to derive a scalar model under a single shear assumption as represented by the top and bottom arrows. Displacements $u$ of the mesh points (represented by the circles) are assumed to be restricted to the horizontal direction. Elastic and plastic interactions are defined among nearest neighbor lattice points only.
	\label{Fig_apendice} }
\end{figure}

In addition to the heuristic presentation given in the main text, and the derivation from a full tensorial model given in \cite{Jagla_Yiel}, we present here an alternative  derivation of the model in a case in which the deformations in the material are assumed from the beginning to be scalar.
This derivation may be applied to a case in which the material is submitted to an external single shear stress (instead of the deviatoric stress $e$ assumed in the main text, which is composed of two orthogonal shears). Under these conditions (see Fig. \ref{Fig_apendice}) we will make the assumption that the local displacements ${\bf u} \equiv (u_x,u_y)$ describing the sample deformation occur only along the $x$ direction, namely $u_y\equiv 0$. We will refer to $u_x$ simply as $u$, and we will use two subindexes to indicate spatial positions in the sample along $x$ and $y$ directions. 

We assume that the interaction between different $u_{ij}$ is local, i.e., it depends on the difference of $u_{ij}$ on neighbor sites.
Along $y$, the interaction between $u_{i,j}$ and $u_{i,j+1}$ must allow for ``slips" between consecutive planes. 
We introduce corrugated potential energy functions $V(u_{i,j+1}-u_{i,j})$ with these properties. 
Along $x$ the interaction between $u_{i,j}$ and $u_{i+1,j}$ is taken to be perfectly harmonic, i.e., described by an energy term $\mu(u_{i+1,j}-u_{i,j})^2/2$. However, and in order to get a final version as similar as possible to the one presented in the main text, we also add the corrugated potential along the $x$ direction with a term $V(u_{i+1,j}-u_{i,j})$. This will not affect qualitatively the phenomenology, since different wells of the corrugated potential along $x$ will not be explored because of the existence of the harmonic term.

The actual form of $V$ is stochastic and depends on the actual values of $i$ and $j$, but we do not indicate this in the notation. The elastic energy of the system is 

\begin{equation}
E=\sum_{ij}  \frac \mu 2 (u_{i+1,j}-u_{i,j})^2 + V(u_{i,j+1}-u_{i,j})+ V(u_{i+1,j}-u_{i,j})
\end{equation}
We will write a first order evolution dynamics for the model. One may think to write something like

\begin{equation}
\dot u_{ij}= -\frac {\delta E}{\delta u_{ij}}
\label{no}
\end{equation}
However, this equation relates the velocity of variation of $u_{ij}$ to the total force acting on $u_{ij}$, and breaks Galilean invariance. It is more natural to postulate a dissipation mechanism in which viscous forces appear when there are {\em relative} motions between neighbor particles. In a mechanical analogy, instead of adding a dashpot between position $u_{ij}$ and a reference position as Eq. (\ref{no}) implies, we add dashpots between neighbor sites on the sample. This leads to write the force balance at position $i,j$ as
\begin{eqnarray}
4\dot u_{i,j}-\dot u_{i,j+1}-\dot u_{i,j-1}-\dot u_{i+1,j}-\dot u_{i-1,j}=\nonumber \\
=\mu (u_{i+1,j}+u_{i-1,j}-2u_{i,j})+\nonumber\\
+F(u_{i,j+1}-u_{i,j})+F(u_{i,j-1}-u_{i,j}) +\nonumber\\
+F(u_{i+1,j}-u_{i,j})+F(u_{i-1,j}-u_{i,j}) 
\end{eqnarray}
where $F(x)\equiv -\partial V(x)/\partial x$.
This is already the model we are seeking for. However, we need to rearrange its terms in order to display its similarity
with the model presented in the text.
We define $e_{i,j}\equiv u_{i,j+1}-u_{i,j}$, in such a way that combining the previous equations at sites $i,j$, and $i,j+1$ we obtain
\begin{eqnarray}
4\dot e_{i,j}-\dot e_{i,j+1}-\dot e_{i,j-1}-\dot e_{i+1,j}-\dot e_{i-1,j}=\nonumber \\
=\mu (e_{i+1,j}+e_{i-1,j}-2e_{i,j})+\nonumber\\
+F(e_{i,j+1})+F(e_{i,j-1})+F(e_{i+1,j})+\nonumber\\
+F(e_{i-1,j})-4F(e_{i,j})
\end{eqnarray}
By introducing the notation 
$\partial^2_x U\equiv U_{i+1,j}+U_{i-1,j}-2U_{i,j}$, and $\partial^2_y U\equiv U_{i,j+1}+U_{i,j-1}-2U_{i,j}$,  
we can write the previous equation in the compact form
\begin{equation}
(\partial^2_x+\partial^2_y)\dot e= -\mu \partial^2_x e+(\partial^2_x+\partial^2_y) F(e) 
\label{a1}
\end{equation}
Note that this equation does not fix the evolution of the mean value $\overline e$, which must be determined according to the driving mechanism that is assumed to hold. %In particular, under a constant strain rate, we must impose $\overline e =\dot \gamma t$. 

Going to Fourier space and dividing by $q_x^2+q_y^2$, Eq. (\ref{a1}) can be written (for ${\bf q}\ne 0$) as
\begin{equation}
\dot e_{\bf q}= -\mu   \frac{q_x^2}{q_x^2+q_y^2}  e_{\bf q} +\left .F(e)\right |_{\bf q}
\end{equation}
In real space this equation reads (introducing the applied stress $\sigma$)
\begin{equation}
\dot e_r=-\frac{dV}{de_r} +\sum _{r'}G_{rr'}e_{r'}+\sigma
\end{equation}
where $G_{r,r}$ is the real space form of 
\begin{equation}
G_{\bf q}\equiv \frac{-\mu q_x^2}{q_x^2+q_y^2} 
\end{equation}
In this form, the structure of the model is seen to be identical to that of Eqs. (\ref{modelo}) and (\ref{gdeq}). Note that the kernel we find here has the property $G_{\bf q}\le 0$, as it was the case for Eq. (\ref{gdeq}). 
The only difference is in the symmetry of $G$, which is now dipolar instead of quadrupolar. This is naturally originated in the single shear geometry assumed in this restricted version.


\begin{references}
\bibitem{footnote} The correlation length exponent $\nu$ is also potential-independent. This follows as a consequence of the statistical tilt symmetry of the problem, that implies $\nu=1/(d-d_f)$, with $d=2$ in our case.
\end{references}
\end{document}